\documentclass[prd,preprint]{revtex4}
\usepackage{tabularx}

\usepackage[dvips]{color}
\usepackage[open]{bookmark}

\newcommand{\h}{\mathcal{H}}
\renewcommand\L{\mathcal{L}}

\newcommand{\AER}[1]{{\color{black} #1}}

	\usepackage{amsmath}%,amssymb} 
	\usepackage{amsfonts}
  	\usepackage{amssymb}
  	\usepackage{float}
	\usepackage{makeidx}
	\usepackage{amsfonts}
	\usepackage[ansinew]{}
	\usepackage[usenames,dvipsnames]{pstricks}
	\usepackage{epsfig}
\usepackage{xcolor}

\newcommand{\be}{\begin{equation}}
\newcommand{\ee}{\end{equation}}
\newcommand{\bea}{\begin{eqnarray}}
\newcommand{\eea}{\end{eqnarray}}

\newcommand\SPD {\nabla^2}

\newcommand\R{\zeta}	
\makeindex

\begin{document}

\preprint{CERN-TH-2023-006}

\title{Effective speed of cosmological perturbations}% Force line breaks with \\
%\thanks{A footnote to the article title}%
\author{Antonio Enea Romano}
\affiliation{Instituto de F\'isica, Universidad de Antioquia, A.A.1226, Medell\'in, Colombia}
%\affiliation{Theoretical Physics Department, CERN, CH-1211 Geneva 23, Switzerland}

\affiliation{ICRANet, Piazza della Repubblica 10, I--65122 Pescara, Italy}
%${}^{3}$Center for Graviational Physics, 
%Yukawa Institute for Theoretical Physics, Kyoto University, Japan \\
%${}^{2}$Department of Physics, University of Crete, 71003 Heraklion, Greece

\date{\today}

\begin{abstract}
We derive an effective equation and action for comoving curvature perturbations and gravitational waves (GWs)  in terms of a time, momentum and polarization dependent effective speed, encoding the effects of the interaction among metric perturbations or with other fields, such as dark energy and dark matter. The structure of the effective actions and equations is the same for scalar and tensor perturbations, and the effective actions can be written as the Klein-Gordon action in terms of an appropriately defined effective metric, dependent on the effective speed. The effective action reproduces, and generalizes to higher order in perturbations, results obtained for GWs in the effective field theory of inflation and dark energy, or for curvature perturbations in systems with multiple scalar fields, encoding in the effective speed  the effects of both entropy and anisotropy. The effective approach can also be applied to the solutions of theories with field equations different from the Einstein equations, by defining an appropriate effective energy-momentum tensor.
As an example, we  show that for a minimally coupled scalar field in general relativity, the effective speeds of curvature perturbations and gravitational waves are frequency and polarization dependent, due to their coupling in the action beyond the quadratic order.
\end{abstract}

\pacs{Valid PACS appear here}% PACS, the Physics and Astronomy
                             % Classification Scheme.
%\keywords{Suggested keywords}%Use showkeys class option if keyword
                              %display desired
\maketitle

%\tableofcontents
% \section{Scalar perturbations}
% \subsection{Frequency dependency of $c_s$ in vanilla inflation}
% \subsection{Example with e.m. and pert EST}
% \subsection{Effective eq derivation from e.m.}

% \section{Tensor perturbations}
% \subsection{Dark matter(bosonic and fermionic) and h}
% \subsection{FRW as effective metric dis-formally transformed}

\section{Introduction}
Cosmological perturbations \cite{Kodama:1985bj} play a fundamental role in modeling the Universe, since they allow to study the origins of large scale structure, the anisotropies of the cosmic microwave background (CMB) radiation, and the propagation of gravitational waves on cosmological distances.  
A momentum dependent effective sound speed (MESS)  for comoving curvature perturbations was defined in \cite{Romano:2018frb}, and applied in \cite{Romano:2020kmj}, showing how the effects of entropy perturbations in multi-fields scalar systems can be encoded in the MESS. Nevertheless that definition of the MESS did not include the effects of anisotropy, and for this reason could not be applied to gravitational waves (GWs). 
In this paper we show how to generalize the effective speed definition in order to encode the effects of any type of interaction or self-interaction, not just entropy perturbations, including anisotropy perturbations, allowing to apply it also to GWs \cite{Romano:2022jeh} and multi-fields systems with anisotropy.
This effective momentum and polarization dependent speed is encoding interaction effects similar to those which make frequency dependent the speed of electromagnetic waves propagating in a medium, inducing birefringence \cite{Romano:2023lxf}, and can be used to model the interaction between cosmological perturbations and dark energy and dark matter.

For cosmological perturbations the effective medium stress-energy tensor  (SET) is given  by the sum of the SET of the  matter fields, the SET of different types of cosmological perturbations, and the SET of the additional fields associated to gravity modification.

We derive a set of  model  independent effective equations and Lagrangians which can describe the evolution of perturbations  in a large class of systems. This includes for example multi-fields systems \cite{Romano:2020kmj}, or modified gravity theories \cite{Vallejo-Pena:2019hgv}. 
Given the generality of this effective description it is particularly suitable for model independent phenomenological analysis of observational data.
The approach predicts naturally that the speed of gravitational waves  can depend on  frequency and polarization, due to the interactions of the graviton with itself or other fields.
These interaction effects could allow to use gravitational waves observations to investigate the elusive nature of dark matter and dark energy.

The equation and Lagrangian for comoving curvature perturbations and  gravitational waves  has the same  structure. The  effects of  interaction can be modeled by a single effective quantity, playing the role of effective propagation speed. This is particularly useful since it allows to compare different models in terms of the  effective speeds of comoving curvature perturbations and gravitational wave, which we denote respectively as $c_s$ and $c_{T,A}$.
Combining different set of observational data such as cosmic microwave background radiation and gravitational waves, it will be possible to constrain $c_s$ and $c_{T,A}$, to determine possible deviations from general relativity and vanilla inflation. 

The effective equations and Lagrangians are derived separately for comoving curvature perturbations and GWs, both in physical and momentum space, and the consistency with, and generalization of  previous results derived in the literature, is considered in different cases.  New effects are then considered, such as the frequency dependency of $c_s$ and $c_{T,A}$ in vanilla inflation or axion inflation, due to the effects of higher order interaction terms.

\section{Effective approach for comoving curvature perturbations}
In this section we will show  that any solution of the EOM equation of comoving curvature perturbations in systems with source terms associated to interaction and self-interaction, can be obtained as a solution of an appropriately defined  effective equation without source terms. This effective approach is conceptually analogous to the description of the effects of the propagation of an electromagnetic wave in a medium in terms of its modified propagation speed.
The method is general, and can be applied to any perturbation, as we will show later for tensor perturbations.
\subsection{Space dependent effective sound speed}
In this section we will show that it is possible to define a space dependent effective sound speed (SESS) in terms of which a model independent equation for comoving curvature perturbations $\zeta$ can be derived.
The SESS encodes the effects of the interaction of $\zeta$ with itself and other fields.

\subsubsection{Effective stress-energy tensor approach}

Scalar perturbations of the metric and of the effective stress-energy tensor (EST) can be written as
\bea
ds^2 = -(1+2A)dt^2+2a\partial_iB dx^idt + a^2\left\{\delta_{ij}(1+2C)+2\partial_i\partial_jE\right\}dx^idx^j \, , \label{pmetric} \\
T^0{}_0 = - (\rho+\delta\rho) \quad \,, \quad  T^0{}_i = (\rho+P) \partial_i(v+B) \,, \nonumber \\ T^i{}_j = (P+\delta P)\delta^i{}_j + \delta^{ik}
\partial_{k}\partial_{j}\Pi
-\frac{1}{3} \delta^{i}{}_{j} \SPD \Pi  \, . \label{psem}
\eea
where $v$ is the velocity potential, $\SPD \equiv \delta^{kl}\partial_k\partial_l$, and $\Pi$ is the anisotropy potential.
Note that any metric and EST can always be written in the above form, making all the results obtained from it completely model independent. 
The comoving slices gauge is defined by the condition $(T^{0}{}_{i})_c=0$, and we  denote with a subscript $c$ quantities evaluated on comoving slices.
In multiple fields systems the EST is the sum of  the stress-energy tensors of each field, which in the comoving gauge is associated to entropy perturbations \cite{Romano:2018frb}.

In the comoving gauge entropy perturbations $\Gamma$ are introduced \cite{Kodama:1985bj} by
\begin{align}
\delta P_c(\eta,x^i) &= c_a(\eta)^2 \delta\rho_c(\eta,x^i) + \Gamma(\eta,x^i) \, , \label{entropy} \end{align}
where $c_a$ is interpreted as the adiabatic sound speed, and is by definition a function of time only.
Note that this definition can be ambiguous \cite{Romano:2018frb}.

The manipulation of the perturbed Einstein's equation in the comoving gauge gives \cite{Naruko:2018fwo}

%\dot{\R} &= - \frac{c_a^2}{a^2H\epsilon} \SPD\Phi_B - \frac{\Gamma}{2H\epsilon} - \frac{1}{3H \epsilon} \SPD \Pi\, , \label{RPhiGamma} \\
\bea
\R''+\frac{\partial_\eta z_a^2}{z_a^2}\R'-c_a^2\SPD\R = a^2\mathcal{S} \,, \label{zetaS} \\
\mathcal{S} =-\frac{c_a^2}{\epsilon}\SPD \Pi- \frac{1}{ 2 a^2 z_a^2 } \left[ \frac{a^3}{c_a^2 H} \left(\Gamma + \frac{2}{3}\SPD \Pi \right) \right]' \label{RS} &,& z_a^2= \frac{\epsilon a^2}{c^2_a} \,,
\eea
where we are denoting derivatives with respect to conformal time with a prime, and $\epsilon$ is the first order slow-roll parameter.
The source term in the above equation can be absorbed into the definition of the space effective sound speed (SESS) following a similar approach to the one adopted for gravitational waves \cite{Romano:2022jeh}.
We will denote with a hat quantities obtained by substituting the solutions of the system of differetial equations describing the full evolution of the physical system,  which are coupled to  eq.(\ref{zetaS}).
We can first re-write eq.(\ref{RS}) as
\be
\frac{(\hat{\R}' z_a^2)'}{z_a^2}-\frac{(g z_a^2)'}{z_a^2}- c_a^2\nabla^2 \hat{\R}=\frac{\Big[\hat{\R}' z_a^2(1-g/\R')\Big]'}{z_a^2}- c_a^2\nabla^2\hat{ \R}=0 \label{gh} \,,
\ee
where we have defined
\be
g=\frac{1}{z_a^2}\int z_a^2 a^2\hat{ \mathcal{S}}\,d\eta\,.
\ee
Note that  after the substitution, $\hat{ \mathcal{S}}(\eta,x^i)$ is a function of space-time coordinates, encoding the integrated effects of the source term on the perturbation propagation.
After introducing the quantities
\bea
1+\delta(\eta,x^i)=\Big(1-\frac{g}{\hat{\R}'}\Big)^{-1/2} \quad &,& \quad \alpha^2=\frac{z_a^2}{(1+\delta)^2}=\frac{\epsilon a^2}{c^2_a(1+\delta)^2} \,, \label{vs}
\eea 
we can rewrite eq.(\ref{gh}) as
\be
\frac{1}{z_a^2}(\alpha^2 \hat{\R}')'-c_a^2\nabla^2 \hat{\R}=0\,. \label{zh}
\ee
Defining the space effective sound speed (SESS) as
\be
c_s(\eta,x^i)=c_a(\eta)\Big[1+\delta(\eta,x^i)\Big] \,, \label{sess}
\ee
and re-writing $\alpha$ in terms of $c_s$ as
\bea
\alpha^2=\frac{\epsilon a^2}{c^2_s} \,,
\eea 
we finally obtain the model independent effective equation
\bea
\zeta''+2 \frac{\alpha'}{\alpha} \zeta'-c_s^2 \nabla^2 \zeta&=&0 \label{zetaalpha} \,,
\eea
which shows that $c_s$ is the correct definition of effective sound speed.
Solving eq.(\ref{zetaalpha}) with the same initial conditions used to obtain $\hat{\zeta}$ will give by construction the same solution. Each different set of initial conditions will give a different SESS. Note that other solutions of eq.(\ref{zetaalpha}), obtained by imposing initial conditions different from those used to compute $\hat{\zeta}$, are not physically relevant. For example $\zeta=0$ is a solution of eq.(\ref{zetaalpha}), but this is not physically relevant, since it corresponds to different initial conditions.

The action corresponding to the effective equation is 

\bea
\mathcal{L}^{eff}_{\zeta}=\frac{z^2}{c_s^2}\Big[\zeta'^2- c^2_s(\nabla \zeta)^2\Big]=\alpha^2\Big[\zeta'^2- c^2_s(\nabla \zeta)^2\Big]
\label{Leffzeta} \quad &,& \quad
\alpha^2=\frac{z^2}{c^2_s}=\frac{\epsilon a^2}{c^2_s} \,,
\eea
Note that in general the effective speed has to be defined from $\mathcal{\hat{S}}$ as shown above, and only in some cases this is equivalent to a manipulation of the Lagrangian, when the EOM simplify to algebraic relations, such as in multi-fields models with heavy fields for example \cite{Achucarro:2012sm}, as shown in  Appendix \ref{ELA}.
Eq.(\ref{zetaalpha}) is general,  and it can be applied to study multi-fields models \cite{Romano:2020kmj,Romano:2018frb}, modified gravity \cite{Vallejo-Pena:2019hgv}, or the interaction with dark energy or dark matter.
We have derived it in the Einstein frame, but a similar approach can also be applied in the Jordan frame. 

Note that the SESS definition given in eq.(\ref{sess}) is more general that the one given in \cite{Romano:2018frb}, which is not including the  effects of  anisotropy.
The generalization presented above, based on absorbing the source term in the effective speed definition, is consistent with the fact that at the Lagrangian level there is no distinction between entropy and anisotropy perturbations, since they are both associated to interaction terms, i.e.  the source terms in the field equations are obtained from the variation of the interaction Lagrangian. 
The SVT decomposition of the EST is valid at any order in perturbations, so eq.(\ref{zetaalpha}) is including the effects of interaction at any order in perturbations, including self-interaction, and for this reason is in agreement with eq.(\ref{zetaalphaA}), obtained by encoding in $c_s$ the effects of all higher order interaction terms. We show this explicitly below.

If  the Einstein equations  (EE) are expanded beyond the linear order,  the effects of the  extra terms can still be encoded in the effective speed by moving them to the r.h.s. and including them in the definition of an effective $\mathcal{S}^{eff}$
\bea
L_1(\hat{\R})+L_2(\hat{\R})&=& a ^2\mathcal{\hat{S}} \,,\label{L12S} \\ L_1(\hat{\R})&=&a^2\mathcal{\hat{S}}-L_2(\hat{\R})=\mathcal{\hat{S}}^{eff}\,,\label{E1N}
\eea
where $\hat{S}$ includes higher order source terms, and $L_1$ is the differential operator corresponding to the l.h.s. of the linear expansion of the  EE, given in the l.h.s. of eq.(\ref{zetaS}), and  a hat denotes quantities obtained by substituting the solution $\hat{\R}$ of eq.(\ref{L12S}), which is including the higher order effects. 
Once the equation has been written in the form given in eq.(\ref{E1N}), it is possible to define the effective speed and effective equation according to the procedure given above, and obtain the solution $\hat{\R}$ of the equation including the effects of the higher order terms as a solution of the effective equation.

% \mathcal{S} =a^2\{\frac{c_s^2}{\epsilon}\SPD \Pi+ \frac{1}{z^2 a} \left[ \frac{a^3}{c_s^2 H} \left(\Gamma + \frac{2}{3}\SPD \Pi \right) \right]'\}&=&0 \,, \label{Rdotcgamma}
% \eea

\subsubsection{Interpretation of the space dependency of the SESS}
The space dependency of the SESS is a natural manifestation of the interaction and self-interaction of  perturbations.
In multiple scalar fields systems these effects are already present in the quadratic Lagrangian, and the source term is  associated to entropy perturbations \cite{Romano:2018frb,Achucarro:2012sm}, while for a single scalar field they manifest only starting from the cubic Lagrangian. 

\subsection{Momentum effective sound speed}
In this section we will show that it is possible to define a momentum dependent effective sound speed (MESS) in terms of which a model independent equation for comoving curvature perturbations $\zeta$ can be derived.
The MESS encodes the effects of the interaction of $\zeta$ with itself and other fields, at any order in perturbations.
The MESS is not the Fourier transform of the SESS, but it is mathematically convenient, since it allows to obtain a model independent equation involving minimal changes of the vanilla case.

\subsubsection{Field equations approach}
Taking the Fourier transform of eq.(\ref{RS}) we get 
\bea
\R_k''+\frac{\partial_\eta z_a^2}{z_a^2}\R_k'+c_a^2 k^2 \R_k = a^2\mathcal{S}_k \,,\label{zetaSk} \\
\mathcal{S}_k =\frac{c_a^2}{\epsilon}k^2 \Pi_k- \frac{1}{2 z_a^2 a^2} \left[ \frac{a^3}{c_a^2 H} \left(\Gamma_k - \frac{2}{3}k^2 \Pi_k \right) \right]' \label{RSk} &,& z_a^2= \epsilon a^2/c^2_a \,,
\eea
After introducing the quantities
\bea
g_k=\frac{1}{z_a^2}\int z_a^2 a^2\mathcal{\hat{S}}_k\,d\eta \quad , \quad
1+\delta_k(\eta)=\Big(1-\frac{g_k}{\hat{\R}_k'}\Big)^{-1/2} \quad , \quad \tilde{\alpha}^2=\frac{z_a^2}{(1+\delta_k)^2}=\frac{\epsilon a^2}{c^2_a(1+\delta_k)^2}\,, \label{vsk}
\eea 
we can rewrite eq.(\ref{RSk}) as
\be
\frac{1}{z_a^2}(\tilde{\alpha}^2 \R_k')'+c_a^2 k^2 \R_k=0\,,\label{zhk}
\ee
where again we are denoting with a hat quantities obtained by substituting the solutions of the system of differential equations describing the full evolution of the system.
Defining the momentum effective sound speed (MESS) as
\be
\tilde{c}_s(\eta,k)=c_a(\eta)\Big[1+\delta_k(\eta)\Big] \,, \label{messa}
\ee
and re-writing $\tilde{\alpha}$ in terms of $\tilde{c}_s$ as
\bea
\tilde{\alpha}^2=\frac{\epsilon a^2}{\tilde{c}^2_s} \,,
\eea 
we finally obtain the model independent effective equation
\bea
\zeta_k''+2 \frac{\tilde{\alpha}'}{\tilde{\alpha}} \zeta_k'+\tilde{c}_s^2 k^2 \zeta_k&=&0 \label{zetaalphak2} \,,
\eea
which shows that $\tilde{c}_s$ is the correct definition of momentum effective sound speed, in agreement with eq.(\ref{zetaalpha}).
\begin{table}[h]

\begin{tabular}{ | m{1.5cm} | m{7cm}| m{8cm} | }
\hline
%Space effective speed  \\ \hline
  & Gravitational waves & Curvature perturbations   \\ \hline
 Speed  & $c^2_{T,A}(\eta,x^i)=\Big(1+\frac{L_h^{int}}{h'^2_A}\Big)^{-1} =\Big(1-\frac{g_A}{h'_A}\Big)^{-1}$  & $c_s^2(\eta,x^i)=\Big(1+\frac{L_{\zeta}^{int}}{\zeta'^2}\Big)^{-1}=c_a^2(\eta)\Big(1-\frac{g}{\R'}\Big)^{-1}$  \\ \hline
 $\mathcal{L}^{eff}$ , $\alpha$  & $\frac{a^2}{c_{T,A}^2}\Big[ h'^2_A-c_{T,A}^2 (\nabla h_A)^2\Big]$ \,,\, $\alpha^2_A=\frac{a^2}{c_{T,A}^2}$& $ \frac{z^2}{c_s^2}\Big[\zeta'^2- c^2_s(\nabla \zeta)^2\Big]$ \,,\, $ \alpha^2=\frac{\epsilon a^2}{c_s^2}=\frac{z^2}{c_s^2}$ \\ \hline
 %$\alpha$ & & $\\ \hline
 Eq. & $h_A''+2 \frac{\alpha_A'}{\alpha_A} h'_A-c_{T,A}^2 \nabla^2 h_A=0$ & $\zeta''+2 \frac{\alpha'}{\alpha} \zeta'-c_s^2 \nabla^2 \zeta=0$ \\ \hline
 Eq. & $h_A''+2 \Big(\frac{a'}{a}-\frac{c'_{T,A}}{c_{T,A}}\Big) h'_A-c_{T,A}^2 \nabla^2 h_A=0$& $\zeta''+2  \Big(\frac{z'}{z}-\frac{c'_s}{ c_s}\Big) \zeta'-c^2_s \nabla^2 \zeta=0$ \\ \hline
\end{tabular}
\caption{The table summarizes the  physical space definition of the effective equation and the effective space dependent speed  of gravitational waves (SEGS) and comoving curvature perturbations (SESS).}
\label{T1}
\end{table}

\subsection{Effective metric description}
In terms of the effective metric 
\be
ds_{eff}^2=\epsilon a^2\Big[c_s d\eta^2-\frac{\delta_{ij}}{c_s}dx^idx^j\Big] \label{geffzeta} \,,
\ee
the effective Lagrangian can be written as
\be
\mathcal{L}^{eff}_{\zeta}=\sqrt{-g}(\partial_{\mu}\zeta \partial^{\mu}\zeta)=g^{\zeta}_{\mu\nu}dx^{\mu}dx^{\nu} \,.
\ee
for which the equation of motion is simply given by the convariant D'Alembert operator

\be
\square \zeta=\frac{1}{\sqrt{-g^{\zeta}}}\partial_{\mu}(\sqrt{-g^{\zeta}}\partial^{\mu}\zeta)=0 \,.
\ee
In this geometrical description the perturbations propagate in an empty curved space, whose geometry is determined by the interaction of the perturbations.
This is conceptually analogous to the general relativistic geometrical interpretation of the effects of gravity in terms of geodesics in a curved space, whose geometry is determined by the EST.
More about this geometrical interpretation will be discussed in a future work.

\subsection{Consistency with previous calculations}
\subsubsection{Minimally coupled scalar field in general relativity}
The vanilla scenario corresponds to $\mathcal{L}^{int}=0$, leading to $\delta=0$, $c_s=c_a=1$.
The quantity $c_w=P'/\rho'$ does not give the correct definition of sound speed, since it does not coincide with  \cite{Romano:2015vxz} the SESS  $c_s=1\neq c_w$.

\subsubsection{K-inflation}
When the interaction Lagrangian is of the form $\mathcal{L}^{int} \propto f(\eta)\zeta'^2$ the SESS is just a function of time $c_s(\eta)=c_a(\eta)$ and $\delta=0$.
In this case the effective action in eq.(\ref{Leffzeta}) and effective eq.(\ref{zetaalphaA}) are in agreement with \cite{Garriga:1999vw}

\be
\mathcal{L}^{eff}_{\zeta}=\frac{z^2}{c_s(\eta)}\Big[\zeta'^2- c^2_s(\eta)(\nabla \zeta)^2\Big]\,.
\label{LeffzetaK}
\ee

\subsubsection{Ultra-slow roll inflation and its generalizations}
Ultra-slow roll inflation (USR) is a particular case of globally adiabatic system, in general characterized by  the vanishing of $\delta P_{nad}=\delta P_{ud}=\delta P-c_w\delta\rho$ on any scale \cite{Romano:2015vxz}, where $c_w=P'/\rho'$, and the subscript "ud" stands for uniform density gauge, defined by the condition $\delta\rho=0$. In USR models the quantity $c_w=P/\rho'$ coincides with  \cite{Romano:2015vxz} the SESS  $c_s=c_w=1$. In other globally adiabatic models such as generalized USR and Lambert inflation \cite{Romano:2016gop} $c_s=c_w\neq1$.
% To be more explicit, the effective action can be obtained by using $c^2_s \SPD\zeta=\SPD\zeta+f(\eta)\zeta'^2$.
\subsubsection{MESS with multiple scalar fields}
The momentum dependency of the sound speed has been found in some specific multi-fields systems \cite{Achucarro:2012sm} where entropy modes can be integrated out analytically. This was generalized in a model independent framework in\cite{Romano:2018frb,Romano:2020kmj}, defining the MESS for an arbitrary multi-field system, including those in which entropy modes cannot be easily integrated out analytically, and for an arbitrary field space metric.

\subsubsection{MESS in modified gravity}
In modified gravity theories an effective entropy and anisotropy can arise in the comoving gauge perturbed field equations, leading to a MESS depending on the specific gravity theory \cite{Vallejo-Pena:2019hgv}.
%An equivalent definition can be obtained from the Lagrangian describing the perturbations, using the effective action approach outlined in the previous section. 

\subsubsection{Effective sound speed and entropy perturbations}

The SESS was introduced for the first time in a model independent way in \cite{Romano:2018frb} as $c^2_s=\delta P_c/\delta \rho_c$, but that definition is only valid for systems with entropy perturbations, but no anisotropy, while the correct generalization including anisotropy was given in this paper in eq.(\ref{sess}).

It is easy to check   that in absence of anisotropy eq.(\ref{sess}) is in agreement with eq.(29) in \cite{Romano:2018frb}. From the perturbations equation we can in fact obtain $\mathcal{S}$, and the corresponding interaction Lagrangian $L_{int}$

\bea
\mathcal{S}&=&-\frac{1}{ 2 a^2 z_a^2 }  \Big(\frac{a^3}{c_a^2 H} \Gamma\Big)'\,,\\  
g&=&\frac{1}{z_a^2}\int z_a^2 a^2 \mathcal{S}\,d\eta\,=-\frac{a\Gamma}{2\epsilon H} \,,\\
c_s^2&=&c_a^2 \Big(1-\frac{g}{\R'}\Big)^{-1}=\Big(1+\frac{a\Gamma}{2\epsilon H \R'}\Big)^{-1} \,,
\eea
corresponding to the Lagrangian
\bea
\L_{int}&=&z_a^2 L_{int}=z_a^2 \frac{a}{\epsilon H}\Gamma\R'=\frac{a^3}{c^2_a H}\Gamma\R' \,, 
\eea
 Note that, as discussed in the Appendix \ref{ELA}, in general the Lagrangian and EOM definition of the effective speed do not coincide, and the correct one is the one given in terms of the EOM.

This shows explicitly that entropy and curvature perturbations are coupled already at second order in the term $L_{int} \propto \Gamma\R'$, explaining why a momentum dependent $c_s$ arises already from the quadratic action, while the calculation of the momentum dependency due to the anisotropy require the cubic action.

\subsection{New predictions and applications}

\subsubsection{MESS in vanilla inflation due to self interaction}
Even in the vanilla scenario higher order interaction terms are expected to induce a momentum dependency of the effective sound speed, associated to cubic and higher order terms.
These effects are ignored in leading order calculations, but arise naturally at higher order.

For example, for the scalar perturbations we have the interaction Lagrangian \cite{Maldacena:2002vr}
\begin{align}
\mathcal{L}^{(3)}_{int} =  ~a^4 \left[ \frac{\epsilon^2}{a^2}  \zeta'^2 \zeta + \frac{1}{a^2} \epsilon^2 (\partial_i \zeta)^2 \zeta   \right. 
\left. - 2 \frac{\epsilon}{a}  \zeta' \partial_i \zeta \partial_i \chi -\frac{1}{2} \frac{\epsilon^3}{a^2} \zeta'^2 \zeta + \frac{1}{2} \epsilon  \zeta (\partial_i \partial_j \chi)^2 + \frac{1}{2} \frac{\epsilon}{a^2} \eta' \zeta' \zeta^2 \right], \label{S3}
\end{align}
where $\partial^2\chi=\zeta' \epsilon/a $.
Let's denote with $L_g$ the differential operator corresponding to the Lagrange equations.
The source term in the EOM will be given by $\mathcal{S}=-L_g(\mathcal{L}^{(3)}_{int})$. After solving the EOM the effective speed will be given by the general definitions in eq.(\ref{sess}) and eq.(\ref{messa}).
The MESS encodes the effects of self-interaction on $\zeta$, which are associated to loop corrections of the power spectrum \cite{Kristiano:2022maq}, which can become large when slow-roll is violated \cite{Romano:2016gop}.

\subsubsection{MESS in modified gravity}
For a specific case of Horndeski theory \cite{Vallejo-Pena:2019hgv} the MESS was computed in the comoving gauge, showing explicitly that it is momentum dependent, as expected.
The same result can be extended to other modified gravity theories once the cubic and higher order actions have been computed.
For example for Horndeski theory the cubic action was computed in \cite{Gao:2012ib}, including the coupling of tensor and scalar perturbations.

%When only one scalar field is present the momentum dependency arises only starting from the cubic action because there is no entropy, and only anisotropy can act a a source term, which corresponds to cubic terms in the action.
% ??comoving-unitary gauge general relation in terms of Gamma?  is there any entropy in Horndeski or just anisotropy?

The model independent approach we have derived does not require any definition of entropy perturbations, and it includes the effects of anisotropy, since they are both related to interactions terms.
\subsubsection{f(R) theories}
In the Einstein's frame  $f(R)$ theories are mathematically equivalent to general relativity with a minimally coupled scalar field. The anisotropic part of the EST of the scalar field arises only at second order in scalar perturbations, is proportional to the space derivatives $\delta\phi_{,i},\delta\phi_{,j}$ \cite{Antusch:2016con}, and is associated to cubic terms in the action \cite{Gao:2012ib}, coupling  tensor and scalar perturbations, which are not included in the quadratic actions.% used in \cite{Kobayashi:2011nu}.

The source terms corresponding to the cubic order action  \cite{Maldacena:2002vr} will induce a  momentum dependency of the effective speed predicted by the MESS approach, which can be interpreted as the effect of the anisotropy of the EST, corresponding to cubic self-interaction terms in the Lagrangian, and to the cubic terms coupling  scalar and tensor perturbations. 
% \subsubsection{Einstein vs Jordan frame}
\subsubsection{MESS in axion inflation}
The coupling of scalar perturbations with a gauge field induces a momentum dependency of the MESS, which should arise already in the quadratic Lagrangian.
For example the quadratic interaction Lagrangian can contain terms of the form
\be
L^{(2)int}_\zeta \propto \delta A_{\mu}\partial^{\mu}\zeta \, , \, \delta A_{\mu}\partial^{\mu}h
\ee
where $\delta A_{\mu}$ denotes perturbations of the gauge field $A_{\mu}$, while at higher order other terms can appear such as for example

\be
L^{(3)int}_\zeta \supset  \,\partial_{\nu}(\delta A_{\mu})\partial^{\mu}\zeta\partial^{\nu}\zeta  \quad , \quad \delta A_{\mu}\delta A^{\mu}\zeta \quad , \quad \delta F_{\mu\nu}\partial^{\mu}\zeta\partial^{\nu}\zeta \,,
\ee
where $F_{\mu\nu}$ denoted the perturbations of the Faraday tensor $F_{\mu\nu}=\partial_{\mu}A_{\nu}-\partial_{\nu}A_{\mu}$.
The effects of these interaction terms are often ignored in the literature, but a priori there is no no general argument to justify that they can be always neglected. These interaction terms give rise to effects similar to those associated to entropy in multi-fields scalar systems, which can be important \cite{Romano:2020kmj}, and are need to  be studied systematically.

% Regarding the relation between the friction term and the effective speed, it should be noted that eq.(\ref{zetaalphaA}) is derived in the Einstein's frame, which is related by a conformal transformation to the frame used in \cite{Kobayashi:2011nu}. In the Einstein's frame the only independent quantity is the effective speed, since the effective Planck mass is not independent of $c_T$.
% We can transform the effective action from the Einstein to the Jordan frame by performing a conformal trasnformation
\section{Gravitational waves}
\subsection{Space dependent effective gravitational wave speed}
Adopting an approach similar to the one used for scalar perturbations, we derive an effective action and propagation equation for gravitational waves.
\subsubsection{Effective stress-energy tensor approach}

The perturbed field equations for tensor modes is \cite{Kodama:1985bj} 
\be
h_A''+2 \h h_A'+ \nabla^2 h_A= a^2 \Pi^{eff}_A\label{hEI} \,,
\ee
where $\Pi^{eff}_A$ is the transverse traceless anisotropic part of the energy-momentum tensor, which can be manipulated to get the space dependent effective GW speed  (SEGS) defined in terms of the EST as \cite{Romano:2022jeh}

\be
c_{T,A}^2(\eta,x^i)=\Big(1-\frac{g_A}{\hat{h}'_A}\Big)^{-1} \quad,\quad 
g_A=\frac{1}{a^2}\int a^4 \hat{\Pi}^{eff}_A\,d\eta\,.
\ee
and the effective equation
\be
h_A''+2 \frac{\alpha_A'}{\alpha_A} h'_A-c_{T,A}^2 \nabla^2 h_A=0 \quad,\quad \alpha^2_A=\frac{a^2}{c_{T,A}^2} \,.
\ee
The corresponding effective action is
\be
 \mathcal{L}^{eff}=\frac{a^2}{c_{T,A}^2}\Big[ h'^2_A-c_{T,A}^2 (\nabla h_A)^2\Big] \label{Lheff}\,,
\ee
in agreement with \cite{Romano:2022jeh}, and generalizing, the action obtained using effective field theory in the Einstein frame \cite{Creminelli:2014wna}. 
For more details about the relationship between Einstein and Jordan frame effective action see \cite{Romano:2023ozy}.

\subsection{Momentum effective gravitational wave speed}
Using a method similar to the one used in physical space, it is possible to derive a model independent effective action and equation in momentum space.
The  results are summarized in the table. We are denoting with $\tilde{h}$ the Fourier transform of $h$.

\vspace{1cm}
\begin{table}[]
\begin{tabular}{ | m{1.5cm} | m{7cm}| m{8cm} | }
\hline
 %\multicolumn{3}{c}{Momentum space} \\ \hline
%Space effective speed  \\ \hline
  & Gravitational waves & Curvature perturbations   \\ \hline
 Speed  & $\tilde{c}^2_{T,A}(\eta,k)=\Big(1+\frac{L_{\tilde{h}}^{int}}{\tilde{h}'^2_A}\Big)^{-1} =\Big(1-\frac{\tilde{g}_A}{\tilde{\tilde{h}}'_A}\Big)^{-1}$  & $\tilde{c}_s^2(\eta,k)=\Big(1+\frac{L_{\zeta_k}^{int}}{\zeta_k'^2}\Big)^{-1}=c_a(\eta)^2\Big(1-\frac{\tilde{g}}{\R_k'}\Big)^{-1}$  \\ \hline
 $\mathcal{L}^{eff}$ \,, $\alpha$  & $\frac{a^2}{\tilde{c}_{T,A}^2}\Big[ \tilde{h}'^2_A+k^2\tilde{c}_{T,A}^2  \tilde{h}_A^2\Big]$ \,,\, $\tilde{\alpha}^2_A=\frac{a^2}{\tilde{c}_{T,A}^2}$& $ \frac{z^2}{\tilde{c}_s^2}\Big[\zeta_k'^2+k^2 \tilde{c}_s^2 \zeta_k^2\Big]$ \,,\, $ \tilde{\alpha}^2=\frac{\epsilon a^2}{\tilde{c}_s^2}=\frac{z^2}{\tilde{c}_s^2}$ \\ \hline
 %$\alpha$ & & $\\ \hline
 Eq. & $\tilde{h}_A''+2 \frac{\tilde{\alpha}_A'}{\tilde{\alpha}_A} \tilde{h}'_A+k^2 \tilde{c}_{T,A}^2 \tilde{h}^2_A=0$ & $\zeta_k''+2 \frac{\tilde{\alpha}'}{\tilde{\alpha}} \zeta_k'+k^2 \tilde{c}_s^2 \zeta_k^2=0$ \\ \hline
 Eq. & $\tilde{h}_A''+2 \Big(\frac{a'}{a}-\frac{\tilde{c}'_{T,A}}{\tilde{c}_{T,A}}\Big) \tilde{h}'_A+k^2 \tilde{c}_{T,A}^2 \tilde{h}^2_A=0$ & $\zeta_k''+2  \Big(\frac{z'}{z}-\frac{\tilde{c}'_s}{ \tilde{c}_s}\Big) \zeta_k'+k^2 \tilde{c}_s^2 \zeta_k^2=0$  \\ \hline
\end{tabular}
\caption{The table summarizes the momentum space definition of the effective equation and effective speed  of gravitational waves (MEGS) and comoving curvature perturbations (MESS).}
\end{table}

%\subsubsection{Effective Lagrangian approach}

%\subsubsection{Effective stress-energy tensor approach}

\subsection{Effective metric description}
Similarly to curvature perturbations, the effective Lagrangian for gravitational waves can be written as
\be
\mathcal{L}^{eff}_h=\sqrt{-g_A}(\partial_{\mu}h_A \partial^{\mu}h_A) \,,
\ee
in terms of the effective metric
    \be
    ds^2_A=a^2\Big[c_{T,A}d\eta^2-\frac{\delta_{ij}}{{c_{T,A}}}dx^idx^j\Big] \label{geffh} \,,
    \ee
for which the GW propagation equation can be written in terms of the covariant d'Alembert operator
\be
\square h_A=\frac{1}{\sqrt{-g_A}}\partial_{\mu}(\sqrt{-g_A}\partial^{\mu}h_A)=0 \,.
\ee
As for scalar perturbations, the effects of the interaction of the graviton can be described as the propagation in a curved space whose metric depends on the SEGS. A similar result can be derived in momentum space.
%This equation shows that the effects of the interaction of the graviton with other fields can be effectively described as the propagation in vacuum through a space with the effective metric given in eq.(\ref{geff}). Note that the above effective  metric is only valid to describe GWs propagation, it is not a modification of the background metric describing the full cosmological model.
\subsection{Consistency with previous calculations}

The model independent  Lagrangians derived in the previous sections are consistent with and extend previous   quadratic action calculations.

\subsubsection{Effective field theory of inflation}
The quadratic order action for tensor modes obtained using the effective field theory of inflation \cite{Creminelli:2014wna} is in agreement with eq.(\ref{Lheff}), but the latter includes also higher order interaction terms neglected in the quadratic action, which induce the polarization and frequency dependency of the speed.

% \subsubsection{Effective field theory of dark energy}
% The quadratic order action derive in \cite{} is in agreement with eq.(\ref{Lheff}) at second order, and g

\subsubsection{Horndeski's theory}

The quadratic action for Horndeski's theory has been computed in \cite{Kobayashi:2011nu,DeFelice:2011bh}.
These calculations are in the Jordan frame,
while in the previous section we have used the Einstein frame. After performing the appropriate conformal transformation \cite{Romano:2022jeh} it can be shown that the tensor modes effective action is in agreement at second order, while eq.(\ref{Lheff}) is generalizing the quadratic action by including  the effects of higher order interaction terms \cite{Gao:2012ib}, associated to self interaction and tensor-scalar coupling.

\subsection{New predictions and applications}
In this section we consider some examples of the application of the effective approach derived previously.

\subsubsection{Scalar tensor feedback and effective speeds}

As an example, let's consider the interaction term

\be
b \,\mathcal{L}^{(3)}_{\R\R h}=b \, a^2  h_{ij}\partial^i\R \partial^j\R= b \, a^2  L^{(3)}_{\R\R h}=z^2 \frac{b}{\epsilon}\, \Big[h_+(\partial^{x}\R \partial^{x}\R - \partial^{y}\R \partial^{y}\R)+ 2 h_{\times}(\partial^{x}\R \partial^{y}\R)\Big]=z^2 \frac{b}{\epsilon} \Big[h_+ \pi_+ + h_{\times} \pi_{\times}\Big] \,,%+2 h_x(\partial^{x}\R\partial^{y}\R) ] 
\ee
which arises at cubic order in general relativity \cite{Maldacena:2002vr} and modified gravity theories \cite{Gao:2012ib}.
The Langrange equations give 
\bea
h_A''+2 \h h_A'+ \nabla^2 h_A&=&  \, a^2 \,b \, \pi_A = a^2 \Pi_A \,, \\
\R''+2  \frac{z'}{z} \R'-c^2_s \nabla^2 \R&=&  \,a^2 \, b \, h^{ij}\partial_i\partial_j \R=a^2 \mathcal{S}\,.
\eea
Contrary to quadratic action,  tensor and scalar modes are coupled at cubic order , and it is necessary to solve a system coupled differential equations to compute the effects of the interaction.
Using the field equations approach the effective sound speed for scalar and tensor modes can be computed.

In terms of the SESS and SEGS, the system of coupled differential equations with source terms reduces to three independent equations, without sources
\bea
h_A''+2 \frac{\alpha_A'}{\alpha_A} h'_A-c_{T,A}^2 \nabla^2 h_A&=&0  \quad,\quad \alpha^2_A=\frac{a^2}{c^2_{T,A}} \,,\\
\zeta''+2 \frac{\alpha'}{\alpha} \zeta'-c_s^2 \nabla^2 \zeta&=&0 \quad,\quad \alpha^2=\frac{\epsilon a^2}{c_s^2}=\frac{z^2}{c_s^2} \,.
\eea

The effects of the interaction induce a modification of both speeds, since the interaction produces a source term in both equations, while in the literature often only the effects on gravitational waves are considered, which are  called scalar induced GWs, ignoring those on scalar perturbations, and their  back-reaction on tensor modes.

\section{Relation to other effective approaches}
The effective approach  formulated in this paper is completely general, and as such includes all the effects of interaction at any order in a single effective quantity, and for an arbitrary number fields.
We can compare this with previous results to see how it includes and extend them.

\subsection{Effective field theory of inflation}

The effective approach we have derived can describe the evolution of curvature  not only for single field models, but also for multi-fields models \cite{Romano:2020kmj}. The EFT of inflation \cite{Cheung:2007st} approach describes elegantly the physics of systems with a single scalar degree of freedom, but when more fields are present, the unitary gauge does not coincide anymore with the comoving gauge \cite{Vallejo-Pena:2019hgv}, and as a consequence entropy and anisotropy perturbations arise \cite{Naruko:2018fwo}, and couple to comoving curvature perturbations. In this  case  the EFT approach based on symmetry breaking does not allow to derive a general effective action  for comoving curvature perturbations \cite{Senatore:2010wk}.

On the contrary the MESS approach  allows to compute the effects of entropy and anisotropy on curvature perturbations for a generic system, including any number of fields, and predicts naturally the momentum dependence of the effective  speed of comoving curvature perturbations.
%??? add effective action from effective equations

%\subsection{Effective Newton constant}

\subsection{Effective field theory of dark energy}

The effective field theory of dark energy \cite{Gubitosi:2012hu} applies the same symmetry breaking idea of the EFT of inflation to dark energy, but in the Jordan frame.
The action is expanded to quadratic order, and for this reason is missing the frequency and polarization dependence of the effective speed which arises naturally in  the MESS and SEGS approach, due to the higher order interactions terms.
The  relation between Jordan and Einstein will be studied in details in a future work.

\AER{
\subsection{Effective field theory of large scale structure}

The effective field theory of large scale structure \cite{Baldauf:2014qfa,Ivanov:2022mrd}  has been developed to study the density contrast $\Delta$, and allows to make predictions about the large scale structure of the Universe which can be tested with galaxy catalogs observations. The comoving density contrast and the Bardeen potential are simply related by the Poisson equation, and the Bardeen potential is related to the comoving curvature perturbations \cite{Romano:2018frb}, so it should be possible to derive an effective equation for $\Delta$ from the one for $\zeta$.
An effective equation for $\Delta$ could be important, since it may allow to understand the relation between the effective speed and the sound horizon in baryonic acoustic oscillations (BAOs). 
We leave the derivation of an effective equation for the density contrast to a future work. 
}

\section{Quantum field theory implications}
The effective Lagrangians we have derived are based on classical calculations, but  they can  be related \cite{Bonifacio:2022vwa} to  the wavefunction of the scalar and tensor fields by the path integral

\be
\Psi[\varphi] \ =  \hspace{-0.4cm} \int\limits_{\substack{\phi(t) \,=\,\varphi\\ \hspace{-0.45cm}\phi(-\infty)\,=\,0}} 
\hspace{-0.5cm} \raisebox{-.05cm}{ ${\cal D} \phi\, e^{iS[\phi]}\,.$ }
\ee
where $\phi$ is a generic field, which in our case could be $\R$ or $h$, and $S$ denotes the action.
At tree level the path integral can be approximated by the action evaluated on the classical solution, which is the way in which we define the effective Lagrangian.

From the wavefunction we can  compute the equal-time correlators as
\be
\langle\phi_1 \cdots \phi_N\rangle = \int{\cal D}\phi\, \phi_1\cdots\phi_N\left\lvert\Psi[\phi]\right\rvert^2\,.
\ee
%
%which can be expanded perturbatively to give a relation between correlation functions and their corresponding wavefunction coefficients, as we will describe.
The above method should give the same result obtained by using canonical quantization in the  in-in formalism \cite{Maldacena:2002vr}.
Following this method we can for example compute corrections to the spectrum, arising from higher order interaction terms in the Lagrangian, in terms of the effective speed we have defined previously. More details about this approach will be given in a separate work.

\AER{
\section{Model independent phenomenological  data analysis}
We have shown that a given solution of the perturbations equations with source can be obtained as the solution of the effective equation without source. Since the solutions of a large class of theories can be obtained from the effective speed,  it can be treated as a free phenomenological quantity which can be constrained by observations. For comoving curvature perturbations this  approach was applied in \cite{Rodrguez:2020hot}, showing that CMB anomalies can be alleviated by an appropriate momentum dependent effective speed encoding  the effects of entropy perturbations, while in this paper we have given the generalization for both tensor and scalar perturbations, including the effects of a generic source term.

The  advantages of  applying this effective approach to observational data analysis are
\begin{itemize}
    \item A single effective quantity can be used for model independent data analysis
    \item The ratio between the  gravitational and electromagnetic luminosity distance is given by the ratio of the GWs effective speed \cite{Romano:2022jeh,Romano:2023ozy} 
\end{itemize}
For GWs we do not have yet observatories operating at very different frequencies, but when space observatories will be available it will be possible to constrain the momentum and polarization dependence of the GWs effective speed \cite{Romano:2024apw}.
An example of the forecast of the constraints on the GWs effective speed from multiband detection can be found in \cite{Baker:2022eiz}, while constraints on the time dependency of the GWs speed using multimessenger observations were derived in \cite{Romano:2023bge}.

\section{Application to modified gravity theories}
The effective approach we derived assuming Einstein equations, can also be applied to other gravity theories, by appropriately defining an effective ES tensor according to
\bea
F[\hat{g}_{\mu\nu}]&=&T_{\mu\nu}[\hat{g}_{\mu\nu},\hat{\phi}_i] \quad\,,\quad M_i[\hat{\phi}_i]=0\,,\label{Fg}\\
G_{\mu\nu}[\hat{g}_{\mu\nu}]&=&\hat{T}^{eff}_{\mu\nu}\,,\label{Gg} \\
\hat{T}^{eff}_{\mu\nu}(x^{\rho})&=&T_{\mu\nu}[\hat{g}_{\mu\nu},\hat{\phi}_i]-F[\hat{g}_{\mu\nu}]+G_{\mu\nu}[\hat{g}_{\mu\nu}]\,,
\eea
where $F$ and $M_i$ are the differential operator corresponding to the gravity and matter fields equations, $T_{\mu\nu}$ is the matter energy-momentum tensor, $\{\hat{g}_{\mu\nu},\hat{\phi}_i\}$ are solutions of eq.(\ref{Fg}), and  the components of $\hat{T}^{eff}_{\mu\nu}$ are functions of space and time, obtained by substituting into ${T}^{eff}_{\mu\nu}$ the solutions of the coupled  matter and gravity field equations (\ref{Fg}). In the context of cosmological perturbations the metric and the energy-momentum tensor are decomposed as the sum of a background and the perturbation component
\bea
\hat{g}_{\mu\nu}=\hat{g}^{0}_{\mu\nu}+\delta \hat{g}_{\mu\nu} &\quad,\quad&
\hat{T}^{eff}_{\mu\nu}=\hat{T}^{0,eff}_{\mu\nu}+\delta \hat{T}^{eff}_{\mu\nu} \,,
\eea
which substituted in eq.(\ref{Gg}) give the perturbed equations in the Einstein-like form 
\be
\delta \hat{G}_{\mu\nu}=\delta \hat{T}^{eff}_{\mu\nu} \,.
\ee
The canonical form of the perturbations equations with source term can be obtained also by manipulating directly the perturbations equations, as shown for gravitational waves in \cite{Romano:2024apw}.
A similar  manipulation of the field equations is used for example in modified gravity  theories \cite{Bellini:2014fua} to obtain an effective Friedman equation in terms of appropriately defined effective dark energy density and pressure, which is an application to the Friedman solution of the general procedure outlined above. 

The above equations should be interpreted as a statement that a given solution $\hat{g}_{\mu\nu}$ of eq.(\ref{Fg}) can also be obtained as a solution of eq.(\ref{Gg}), for an appropriate choice of $\hat{T}^{eff}_{\mu\nu}$ and boundary conditions, not as a statement about the full equivalence between  general relativity and the generic theory corresponding to eq.(\ref{Fg}), which may indeed involve different differential operators, and be fundamentally different. From a phenomenological point of view
eq.(\ref{Gg}) is enough to obtain an effective description of the solutions of eq.(\ref{Fg}). This implies that both the background and the perturbations solutions for theories with field equations different from the Einstein equations, can be obtained as solutions of the Einstein equations with appropriately defined background and perturbed energy-momentum tensors, and consequently the effective method we developed can be applied to obtain an effective description of the solutions of the cosmological perturbation equations of these theories.
Using the method outlined above it can be interesting to compute the effective speed of cosmological perturbations in different modified gravity theories, for example Horndeski theories with matter interaction \cite{Moretti:2020kpp,Moretti:2021ljj}. We leave  detailed model specific calculations to a future work, but we give below a generic example to outline the general mathematical method. 

Let's consider the case of a wave equation  in Minkowski background 
\be
\zeta''+k^2\zeta=\Pi \,,\label{zetaP}
\ee
and compute the effective speed corresponding to dumped oscillatory solutions of the form
\be
\hat{\zeta}=A \,e^{-\lambda t }\,e^{i \omega t} \,,
\ee
where $\lambda$ is the dumping factor, which we introduce as a phenomenological parameter,  motivated by the Landau damping effect obtained in some modified gravity models \cite{Moretti:2020kpp,Moretti:2021ljj}.
The corresponding $\hat{\Pi}$ is obtained by substituting $\hat{\zeta}$ in the l.h.s. of eq.(\ref{zetaP})
\be
\hat{\Pi}=A \left[k^2+(\lambda -i \omega )^2\right] e^{-\lambda  t+i t \omega } \,,
\ee
since this  implies  
\be
\hat{\zeta}''+k^2\hat{\zeta}=\hat{\Pi} \, \,,\label{zetaPhat}
\ee
i.e. $\hat{\zeta}$ is by construction a solution of the equation with source given above. Note that we are assuming that $\hat{\zeta}$ has been obtained independently by solving the full system of differential equations corresponding to the given system, and that in general these equations could have a form different from eq.(\ref{zetaP}), but according to the procedure outlined above the function $\hat{\zeta}$ also satisfies eq.(\ref{zetaPhat}), i.e. we can always obtain an equation having the canonical form and admitting by construction the same solution.
We can now compute the effective speed
\bea
\hat{g}(t)&=&\int \hat{\Pi} \,d\,t=\frac{A\left(k^2+(i \omega- \lambda )^2\right) e^{-\lambda  t+i t \omega }}{i \omega-\lambda } \,, \\
\hat{c}^2(k)&=&\Big(1-\frac{\hat{g}}{\hat{\R}'}\Big)^{-1}=-\frac{(\lambda -i \omega )^2}{k^2} \,, \label{cdump}
\eea
leading to the effective equation
\be
\zeta''+\hat{c}^2(k) k^2\zeta=\zeta''-(\lambda -i \omega )^2 \zeta=0\,,\label{zetaPP}
\ee
which is easy to check  admits $\hat{\zeta}$ as a solution. In this case the effective speed does not depend on time, so there is no friction term, but in general the speed could depend on time, and a friction term be present.
We have shown that the dumped solution of the equation with source can be obtained as a solution of the effective equation without source for an appropriate  definition of the momentum dependent effective speed, given in eq.(\ref{cdump}). In this case the effective speed can be computed analytically, but in general a numerical calculation has to be performed to obtain the effective speed corresponding to a given function $\hat{\zeta}$, giving a general procedure to obtain the effective speed for any solution. The some mathematical procedure can be applied to different cosmological perturbations, for example for each GW polarization  \cite{Romano:2024apw}.

\section{Conclusions}

We have derived a set of  model  independent effective equations and Lagrangians for comoving curvature perturbations and gravitational waves, which can be applied for example to multi-fields systems, or modified gravity. The effective approach has been formulated both in physical and momentum space, clarifying the distinction between the space and momentum dependent effective speeds.
For theories whose fields equation are not of the Einstein-like form we have shown how an appropriate manipulation allows to define an effective energy-momentum tensor, such that the solutions of those theories are also solutions of Einstein-like  equations.
Given the generality of this effective approach it is particularly suitable for model independent phenomenological analysis of observational data. Another advantage is that the ratio between  gravitational and electromagnetic luminosity distance is given by the  effective speed  of GWs, allowing to constrain the GWs effective speed directly with multi-messenger observations.
The effective approach predicts naturally that the speed of gravitational waves can depend on  time, frequency and polarization, due to the interactions of the graviton with itself or other fields.
These interaction effects could allow to use gravitational waves observations to investigate the interaction of GWs with dark matter and dark energy.

The equation and Lagrangian for scalar and tensor perturbations has the same  structure, and the effects of the interaction can be modeled at any order in perturbations by a single effective quantity, playing the role of effective propagation speed. This is particularly useful since it allows to compare different models in terms of the two quantities $c_s$ and $c_T$.
Combining different sets of observational data such as cosmic microwave background radiation and gravitational waves, it will be possible to constrain  $c_s$ and $c_{T,A}$, to determine possible deviations from general relativity and vanilla inflation. If a deviation is found, the theoretical research can be focused on those models able to predict the $c_s$ and $c_T$ supported by observations.

In this paper we have focused on gravitational waves and comoving curvature perturbations, but the density contrast is also very important, since it allows to make predictions about the large scale structure of the Universe, which can be verified with galaxy catalog observations. In the future it will be interesting to apply the effective speed approach to derive an effective equation for the density contrast, and to investigate the relation between the effective speed of the density contrast and the sound horizon in baryonic acoustic oscillations (BAOs).
In order to test specific models it will be important to perform higher order perturbations calculations for different models, in order to compute the effects which are not included in the quadratic action, and in the EST approach are treated as effective model independent phenomenological quantities. 
}

\begin{acknowledgments}
\numberwithin{equation}{section}
I thank   Sergio Vallejo, Riccardo Sturani,  Rogerio Rosenfeld,  Alexander Vikman,  Filippo Vernizzi, Francesco Pace, Antonio De Felice, Shinji Mukohyama, and Antonaldo Diaferio for interesting discussions. I thank  CERN theory division, ICTP-SAIFR, YITP, and Osaka University Theoretical Astrophysics Group for the kind hospitality during the preparation and revision of this paper. This work was supported by the UDEA projects  2021-44670,  2019-28270, 2023-63330.
\end{acknowledgments}

\appendix

\section{Effective Lagrangian approach}
\label{ELA}
We have shown that the effective speed can always be obtained from the solutions of the EOM equations, and an effective action can be obtained from the effective equation. In some cases it is possible to obtain the same action by substituting the solutions of the EOM in the action.
Note that in general this is not possible  \cite{Pons:2009ch}, and so the correct general approach is the one based on the equations of motions. Nevertheless we report about it in this appendix to show the connection with other derivations of a single effective theory in  systems where the physics of the problem reduces some EOM to algebraic relations between the fields, which can then be safely substituted in the action \cite{Achucarro:2012sm}.

\subsection{Space dependent effective speed}
In the action approach the EST on the r.h.s. of the Einstein's equations originates from the interaction of scalar perturbations with themselves or other fields. Based on this we can obtain an effective Lagrangian corresponding to the perturbed field equations by introducing  higher order   interaction terms as $\mathcal{L}_{int}(\zeta,\phi^i)$, where $\phi^i$ denotes abstractly all the other fields $\zeta$ is coupled to.

We will use conformal time, and adopt the following notation for the Lagrangian density $L$
\be
S=\int d\eta dx \mathcal{L}=\int d\eta dx z^2 L \quad , \quad  \mathcal{L}=\ z^2 {L}  \quad , \quad
 z^2=\epsilon  a^2 \,,
\ee
where we are assuming a Friedman background (FRW) background with scale factor $a(\eta)$, and $\epsilon$ denotes the first order slow-roll parameter, defined in terms of cosmic time $dt=a^{-1} d\eta$. 

In general relativity the quadratic Lagrangian of the comoving curvature perturbations of a minimally coupled scalar field is 
\bea
\mathcal{L}^{(2)}_{\zeta}=z^2\Big[\zeta'^2-(\nabla \zeta)^2\Big] &,& z^2=\epsilon  a^2\label{vani}.
\eea
We will call the above model vanilla inflation.

Including higher order interaction and self-interaction terms, we can write a general model independent Lagrangian

\bea
\mathcal{L}_{\zeta}=\mathcal{L}^{(2)}_{\zeta}+\mathcal{L}^{int}_{\zeta}= z^2\Big[\zeta'^2- (\nabla \zeta)^2+L^{int}_\zeta(\zeta,\phi^i)\Big]=z^2\Big[\zeta'^2\Big(1+\frac{L^{int}_\zeta}{\zeta'^2}\Big)- (\nabla \zeta)^2\Big] \label{LEint} \,,
\eea
where $\phi^i$ denotes collectively any other field.
From the above equation we can obtain the effective Lagrangian
\bea
\mathcal{L}^{eff}_{\zeta}=\frac{z^2}{c_s^2}\Big[\zeta'^2- c^2_s(\nabla \zeta)^2\Big]=\alpha^2\Big[\zeta'^2- c^2_s(\nabla \zeta)^2\Big]
\label{LeffzetaA} \quad &,& \quad
\alpha^2=\frac{z^2}{c^2_s}=\frac{\epsilon a^2}{c^2_s} \,,
\eea
where we have defined the space effective  sound speed (SESS) according to
\bea
c_s^2(\eta,x^i)=\Big(1+\frac{\hat{L}_{\zeta}^{int}}{\hat{\zeta}'^2}\Big)^{-1} \label{ctS}  \,
\eea 
where a hat denotes quantities obtained by substituting the solutions of the EOM.

The variation of $\mathcal{L}^{eff}_{\zeta}$ gives the model independent equation
\bea
\zeta''+2 \frac{\alpha'}{\alpha} \zeta'-c_s^2 \nabla^2 \zeta&=&0 \,,\label{zetaalphaA}
\eea
which can be also written as
\bea
\zeta''+2  \Big(\frac{z'}{z}-\frac{c'_s}{ c_s}\Big) \zeta'-c^2_s \nabla^2 \zeta&=&0  \,\label{zetacs}\,.
\eea
Note that in deriving the equations of motion for $\zeta$ the SESS has been treated as a function independent of $\zeta$, since the SESS is an effective quantity determined by substituting the solutions of the full theory, including the effects of interaction, into the interaction Lagrangian $L^{int}$.
For any system with a well defined Lagrangian it should be always possible to solve the  equations of motions, but the substitution back into the action is not always possible \cite{Pons:2009ch}, so this definition of the SESS is only valid in some cases, such as in multi-fields modes with heavy fields \cite{Achucarro:2012sm}.

Eq.(\ref{zetaalphaA})  and eq.(\ref{Leffzeta}) show that $\alpha(\eta,x^i)$ can be interpreted as an effective space dependent scale factor, while eq.(\ref{zetacs}) shows explicitly the modification of the friction term induced by the SESS.
Since  these equations are model independent, we can immediately conclude that the friction term cannot be modified if $c_s'=0$. 

The effective Lagrangian can be obtained from the vanilla case in eq.(\ref{vani})
by performing the transformation 
\be
z^2 \rightarrow \alpha^2=\frac{z^2}{c^2_s} \quad,\quad c\rightarrow c_s \,,\label{trans}
\ee
where we are denoting with $c$ the unity sound speed, to avoid ambiguity.
This is in agreement with eq.(\ref{zetaalphaA}), which shows that $\alpha$ can be regarded as an effective scale factor.

The main advantage of using eq.(\ref{zetaalphaA}) is that it is completely  model independent, allowing to study in a systematic way deviations from general relativity or the effects of the interaction with different fields using a single function.
Note that since $\mathcal{L}_{int}$ can be space dependent, also $c_s(\eta,x^i)$ depends on space, with the exception of $\mathcal{L}_{int}\propto f(\eta)\zeta'^2$, when it is only time dependent, which corresponds to K-inflation \cite{Garriga:1999vw}.

In presence of multiple scalar fields the space dependence is manifested already in the quadratic Lagrangian \cite{Vallejo-Pena:2019hgv},  while the effects of anisotropic perturbations requires at least a cubic Lagrangian, because scalar fields anisotropies appear at second order in the EST.

\subsection{Momentum dependent effective speed}
A model independent effective equation and Lagrangian can also be derived in momentum space.

The Lagrangian in momentum space can be written as

\bea
\mathcal{L}_{\zeta_k}=\mathcal{L}^{(2)}_{\zeta_k}+\mathcal{L}^{int}_{\zeta_k}= z^2\Big[\zeta'^2_k+k^2 \zeta_k^2+L^{int}_{\zeta_k}(\zeta_k,\phi_k^i)\Big] \label{Lintk} \,.\\
\eea
The effective Lagrangian is
\bea
\mathcal{L}^{eff}_{\zeta_k}=\tilde{\alpha}^2\Big[\zeta_k'^2+\tilde{c}^2_s k^2 \zeta^2_k\Big]
 &,&
\tilde{\alpha}^2(\eta,k)=\frac{z^2}{\tilde{c}_s^2}=\frac{\epsilon a^2}{\tilde{c}_s^2}\,, \label{Leffzetak}
\eea
where we have defined the momentum effective sound speed (MESS) $\tilde{c}_s$ and effective scalar factor as 
\bea
\tilde{c}_s^2(\eta,k)=\Big(1+\frac{\hat{L}^{int}_k}{\hat{\zeta}'^2_k}\Big)^{-1} \label{csk} &,&
\tilde{\alpha}^2(\eta,k)=\frac{\epsilon a^2}{\tilde{c}_s^2}=\frac{z^2}{\tilde{c}_s^2} \,.
\eea
where a hat denotes quantities obtained by substituting the solutions of the EOM.
The corresponding equation is
\bea
\zeta_k''+2 \frac{\tilde{\alpha}'}{\tilde{\alpha}} \zeta_k'+\tilde{c}_s^2 k^2 \zeta_k&=&0 \,,  \label{zetaalphak} 
\eea
which can be also be written as
\bea
\zeta_k''+2 \Big(\frac{z'}{z}-\frac{\tilde{c_s}'}{\tilde{c}_s}\Big) \zeta_k'+\tilde{c}_s^2 k^2 \zeta_k&=&0 \,. \label{zetacsk}
\eea
Eq.(\ref{zetaalphak})  and eq.(\ref{Leffzetak}) show that $\tilde{\alpha}(\eta,k)$ can be interpreted as an effective momentum dependent scale factor, while eq.(\ref{zetacsk}) shows explicitly the modification of the friction term induced the MESS.
Since  these equations are model independent, we can immediately conclude that the friction terms cannot be modified if $\tilde{c}_s'=0$. 

The effective Lagrangian can be obtained from the vanilla inflation action
\be
\mathcal{L}^{(2)}_{\zeta_k}=z^2\Big[\zeta'^2_k+c^2 k^2\zeta^2_k\Big] \,,
\ee
by the transformation 
\be
z^2 \rightarrow \tilde{\alpha}^2=\frac{z^2}{\tilde{c}^2_s} \quad,\quad c\rightarrow \tilde{c}_s \,,\label{transk}
\ee
where we are denoting with $c$ the unity sound speed, to avoid ambiguity.
This is in agreement with eq.(\ref{zetaalphak}), which shows that $\tilde{\alpha}$ can be regarded as a momentum dependent effective scale factor.
Note that the quantities $\tilde{c_s}$ and $\tilde{\alpha}$ are not the Fourier transform of $c_s$ and $\alpha$.

This definition of the MESS is not general, but can only be applied when the substitution of the solutions of the EOM is possible \cite{Pons:2009ch}.

%\bibliographystyle{h-physrev4.bst}
%\bibliography{Bibliography.bib}
%\end{document}

%\bibliographystyle{h-physrev4.bst}
%\bibliography{Bibliography.bib}

\end{document}